\documentclass[sigconf]{acmart}

\usepackage{algorithm}
\usepackage{algorithmic}
\usepackage{graphicx}  
\usepackage{tabularx}  
\usepackage{tikz}
\usepackage{multirow}
\usepackage{subcaption}
\usetikzlibrary{shapes.geometric, arrows, positioning}

\tikzstyle{process} = [rectangle, minimum width=3cm, minimum height=1cm, text centered, draw=black, fill=blue!30]
\tikzstyle{selection} = [rectangle, minimum width=3cm, minimum height=1cm, text centered, draw=black, fill=green!30]
\tikzstyle{mutation} = [rectangle, minimum width=3cm, minimum height=1cm, text centered, draw=black, fill=red!30]
\tikzstyle{arrow} = [thick,->,>=stealth]
\tikzstyle{data} = [ellipse, minimum width=3cm, minimum height=1cm, text centered, draw=black, fill=yellow!30]
\AtBeginDocument{%
  }

\setcopyright{none}
\acmConference[GECCO 2025 Supplementary paper]{Genetic and Evolutionary Computation Conference}{Accepted}{as Poster}
\copyrightyear{}
\acmYear{}
\acmISBN{}
\acmDOI{}

\begin{document}

\title{PSO-UNet: Particle Swarm-Optimized U-Net Framework for Precise Multimodal Brain Tumor Segmentation}

\author{Shoffan Saifullah}
\authornotemark[1]
\orcid{0000-0001-6799-3834}
\affiliation{%
  \institution{Faculty of Computer Science, AGH University of Krakow}
  \city{Krakow}
  \country{Poland}
  \postcode{30-059}
}
\email{saifulla@agh.edu.pl}
\affiliation{%
  \institution{Department of Informatics, Universitas Pembangunan Nasional Veteran Yogyakarta}
  \city{Yogyakarta}
  \country{Indonesia}
  \postcode{55281}
}
\email{shoffans@upnyk.ac.id}

\author{Rafał Dreżewski}
\orcid{0000-0001-8607-3478}
\affiliation{%
  \institution{Faculty of Computer Science, AGH University of Krakow}
  \city{Krakow}
  \country{Poland}
  \postcode{30-059}}
\email{drezew@agh.edu.pl}

\renewcommand{\shortauthors}{Saifullah et al.}

\begin{abstract}
  Medical image segmentation, particularly for brain tumor analysis, demands precise and computationally efficient models due to the complexity of multimodal MRI datasets and diverse tumor morphologies. This study introduces PSO-UNet, which integrates Particle Swarm Optimization (PSO) with the U-Net architecture for dynamic hyperparameter optimization. Unlike traditional manual tuning or alternative optimization approaches, PSO effectively navigates complex hyperparameter search spaces, explicitly optimizing the number of filters, kernel size, and learning rate. PSO-UNet substantially enhances segmentation performance, achieving Dice Similarity Coefficients (DSC) of 0.9578 and 0.9523 and Intersection over Union (IoU) scores of 0.9194 and 0.9097 on the BraTS 2021 and Figshare datasets, respectively. Moreover, the method reduces computational complexity significantly, utilizing only 7.8 million parameters and executing in approximately 906 seconds, markedly faster than comparable U-Net-based frameworks. These outcomes underscore PSO-UNet’s robust generalization capabilities across diverse MRI modalities and tumor classifications, emphasizing its clinical potential and clear advantages over conventional hyperparameter tuning methods. Future research will explore hybrid optimization strategies and validate the framework against other bio-inspired algorithms to enhance its robustness and scalability.
\end{abstract}

\ccsdesc{Computing methodologies~Image processing}
\ccsdesc{Computing methodologies~Image segmentation}
\ccsdesc{Networks~Network architectures}
\ccsdesc{Theory of computation~Bio-inspired optimization}

\keywords{Medical Image Segmentation, Particle Swarm Optimization (PSO), U-Net Architecture, Brain MRI Images, Computational Efficiency.}

\received{18 January 2025}
\received[revised]{19 March 2025}
\received[accepted]{19 March 2025}

\maketitle

\section{Introduction}
\label{introduction}
Medical image segmentation is crucial for accurate diagnosis, treatment planning, and patient monitoring. In brain tumor analysis, Magnetic Resonance Imaging (MRI) provides detailed insights but presents challenges due to tumor variability, noise, and inconsistencies in multimodal datasets~\cite{Wadhwa2019,Saifullah2025_mtap}. The BraTS dataset (T1, T2, FLAIR, T1Gd modalities) offers complementary tissue characteristics, such as FLAIR for tumor boundaries and T1Gd for vascular contrast~\cite{Ding2024}. However, integrating these features remains a complex task~\cite{Chukwujindu2024}. Similarly, multi-class datasets like Figshare, with diverse tumor morphologies (e.g., meningioma, glioma, pituitary), introduce additional challenges in segmentation~\cite{Saifullah2024,Saifullah2024_gecco}.

Deep learning architectures, particularly U-Net~\cite{Hernandez-Gutierrez2024,Bouchet2023,Yousef2023}, widely used for medical image segmentation, rely on their encoder-decoder structures with skip connections to capture spatial and contextual information~\cite{Fang2023,Rehman2021}. Despite these advantages, traditional U-Net architectures face critical limitations, notably their performance dependency on hyperparameter tuning~\cite{Asiri2024,Das2022}, such as the number of filters, kernel sizes, and learning rates, which are usually performed manually. This manual tuning is computationally expensive, dataset-specific, and often results in suboptimal performance, hindering their applicability and scalability in clinical settings~\cite{Saifullah2024,Ranjbarzadeh2024}. 

To overcome these limitations, metaheuristic algorithms have emerged as promising solutions for automatic hyperparameter optimization, efficiently navigating complex, high-dimensional search spaces. Particle Swarm Optimization (PSO), in particular, has demonstrated superior capability in balancing exploration and exploitation, rapidly converging to optimal or near-optimal hyperparameter configurations. PSO’s simplicity, computational efficiency, and proven performance in similar optimization scenarios make it highly suitable for integration with deep learning-based segmentation frameworks.

Motivated by these advantages, this study introduces PSO-UNet, a novel integration of PSO and U-Net architecture designed explicitly for multimodal brain tumor segmentation tasks. The primary contributions of this work include:

\begin{itemize}
    \item \textbf{Efficient Hyperparameter Optimization}: Leveraging PSO to dynamically optimize critical U-Net hyperparameters (number of filters, kernel sizes, learning rates), significantly improving segmentation accuracy and efficiency.
    \item \textbf{Enhanced Dataset Adaptability}: Demonstrating robust generalization capabilities across different MRI modalities (FLAIR, T1, T2, T1Gd) and varied tumor types (Meningioma, Glioma, Pituitary).
    \item \textbf{Lightweight and Computationally Efficient Framework}: Reducing the computational burden, achieving state-of-the-art performance with significantly fewer parameters and shorter execution time than traditional U-Net approaches.
    \item \textbf{Rigorous Comprehensive Evaluation}: Benchmarking against leading segmentation methods to establish clear advantages in accuracy, generalization, and efficiency.
\end{itemize}

The remainder of this paper is organized as follows: Section~\ref{related_works} reviews related work on medical image segmentation, focusing on U-Net variants and PSO applications. Section~\ref{method} details the PSO-UNet framework, including its architecture and optimization process. Section~\ref{results-and-discussion} presents experimental results that compare PSO-UNet with existing methods. Finally, Section~\ref{conclusion} concludes with key findings and directions for future research.

\section{Related Works}
\label{related_works}
Medical image segmentation has evolved from traditional handcrafted algorithms to state-of-the-art deep learning methods. Early techniques, such as thresholding, region-growing, and active contour models~\cite{Sheela2020} relied on domain-specific heuristics and predefined rules, which limited their adaptability, robustness, and accuracy, particularly when processing complex, noisy, or multimodal brain MRI datasets~\cite{Liu2024}. These approaches struggled with low-contrast images and anatomical variability, underscoring the need for automated and scalable segmentation methods~\cite{Xu2024}.

Deep learning architectures, specifically U-Net, have revolutionized medical image segmentation due to their robust encoder-decoder structures with skip connections, enabling effective extraction and preservation of spatial and contextual information. Various enhancements to the standard U-Net architecture have been proposed, including ResUNet~\cite{Murmu2023,Ingle2022}, DenseUNet+~\cite{Cetiner2023}, and Attention U-Net~\cite{Bouchet2023,Saifullah2025} have introduced innovations such as residual blocks~\cite{Saifullah2025_ieee}, dense connectivity, and attention mechanisms to improve segmentation accuracy~\cite{Saifullah2024_ijece,Saifullah2024}. Although these variants often outperform traditional approaches, these advancements often come with increased computational complexity and a critical dependence on extensive hyperparameter tuning, which is typically performed manually, resulting in high computational costs, limited scalability, and challenges in clinical settings~\cite{Chukwujindu2024}.

Multimodal and multi-class segmentation tasks introduce additional complexity, as seen in popular datasets such as BraTS~\cite{Baid2023} and Figshare~\cite{Cheng2017}. These datasets incorporate various MRI modalities (e.g., T1, T2, FLAIR, and T1Gd), each modality highlighting different tumor characteristics~\cite{Qiu2025}, requiring sophisticated integration methods to ensure robust generalization across various data types. Similarly, segmentation frameworks must adapt effectively to varied tumor types (e.g., glioma, meningioma, pituitary tumors), each with unique morphologies and spatial complexities~\cite{Saifullah2024_icai3s,Saifullah2023}. Standard U-Net architectures~\cite{Hernandez-Gutierrez2024,KumarBhatt2023,Saifullah2025} often exhibit limited flexibility, resulting in suboptimal performance when faced with this variability and complexity, particularly under conditions of class imbalance or subtle tissue differentiation.

To address these limitations, researchers have explored automated hyperparameter optimization techniques to efficiently navigate high-dimensional search spaces. Prominent optimization strategies such as Bayesian Optimization, Genetic Algorithms (GA), and Differential Evolution (DE) have demonstrated potential to achieve optimal hyperparameter configurations~\cite{Vincent2023}. Bayesian Optimization uses probabilistic modeling for efficient search, but often faces challenges with high computational overhead when dealing with large parameter spaces~\cite{Binois2022}. Genetic Algorithms provide robustness through evolutionary mechanisms but require extensive population initialization and prolonged execution times~\cite{Li2024}. In contrast, particle swarm optimization (PSO) offers a more efficient approach, using collective intelligence to rapidly converge to optimal solutions through balanced exploration and exploitation, thus significantly reducing computational time and complexity~\cite{Gad2022,Priyadarshi2025}.

Recent studies integrating PSO with deep learning, such as UNet-T-PSO~\cite{Saifullah2024}, have demonstrated promising results, showcasing the effectiveness of PSO in tuning hyperparameters for improved accuracy. However, existing research rarely addresses multimodal and multi-class medical image segmentation comprehensively, highlighting a notable research gap. Motivated by these observations, this paper proposes the PSO-UNet framework, explicitly addressing these critical challenges by dynamically optimizing U-Net hyperparameters, thus enhancing segmentation accuracy and computational efficiency, while maintaining robust generalization across multiple MRI modalities and tumor classes.

\section{Method}
\label{method}
This section presents the PSO-UNet framework for medical image segmentation, focusing on its architecture, the integration of Particle Swarm Optimization (PSO) for hyperparameter tuning, and the details of implementation and evaluation. 

\subsection{Overview of the Proposed Framework (PSO-UNet)}
\label{3.1-pso-unet}

\begin{figure}
\centering
\includegraphics[width=0.4\textwidth]{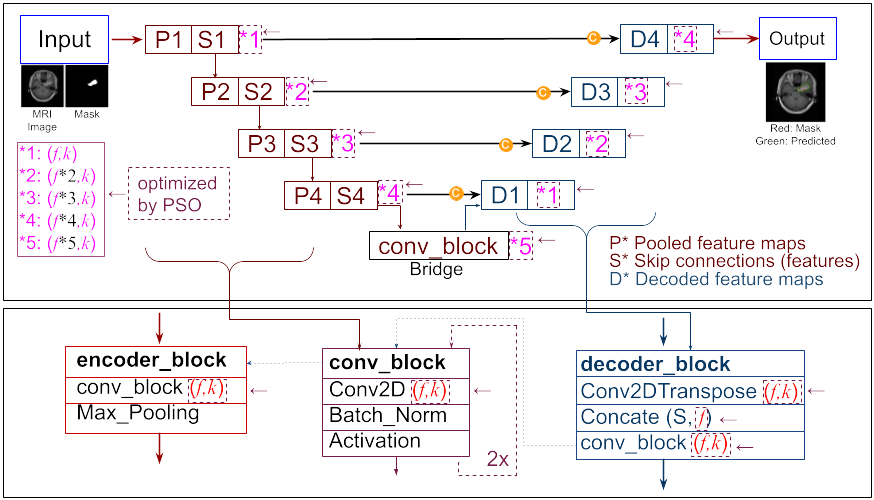}
\caption{U-Net framework optimized by PSO with an encoder-decoder structure and skip connections to enhance spatial and semantic feature retention.}
\label{fig:unet_arch}
\end{figure}

The proposed PSO-UNet framework integrates U-Net feature extraction with PSO optimization to improve the segmentation of multimodal MRI and multi-class tumors. It automates and optimizes critical hyperparameters dynamically, such as filters (\( f \)), kernel size (\( k \)), and learning rate, improving accuracy while ensuring computational efficiency. Figure\ref{fig:unet_arch} explicitly demonstrates how PSO dynamically optimizes these hyperparameters within each U-Net block. The U-Net architecture, known for its encoder-decoder structure and skip connections, effectively captures spatial details and contextual features crucial for precise segmentation tasks~\cite{Saifullah2024,Saifullah2025}. The encoder extracts features through convolutional layers and max-pooling, as described in Eq.\ref{eq_01}.

\begin{equation} \label{eq_01}
\mathbf{F} = \sigma(\mathbf{W} * \mathbf{X} + \mathbf{b})
\end{equation}

Here, $\mathbf{F}$ represents the feature map, $\mathbf{W}$ is the kernel weight, $\mathbf{X}$ is the input feature map, $\mathbf{b}$ is the bias term, and $\sigma$ is the activation function (ReLU). Max-pooling reduces spatial dimensions, enabling the encoder to capture high-level semantic features.

The decoder restores spatial resolution using upsampling layers. Skip connections concatenate encoder features with upsampled decoder features to retain spatial precision, as expressed in Eq.~\ref{eq_02}.

\begin{equation} \label{eq_02}
\mathbf{C} = \mathbf{U} \oplus \mathbf{E}
\end{equation}

Where $\mathbf{C}$ is the concatenated feature map, $\mathbf{U}$ represents upsampled decoder features, and $\mathbf{E}$ represents corresponding encoder features. The segmentation map is generated via a $1 \times 1$ convolution followed by a softmax or sigmoid activation, as shown in Eq.~\ref{eq_03}.

\begin{equation} \label{eq_03}
\mathbf{Y} = \text{Softmax}(\mathbf{W}_{\text{out}} * \mathbf{C} + \mathbf{b}_{\text{out}})
\end{equation}

While U-Net’s encoder-decoder structure enables feature extraction and reconstruction, its performance relies on hyperparameter selection, optimized by PSO in the proposed framework. Manual tuning of these hyperparameters is not only time-consuming but often results in suboptimal configurations, especially across varying MRI modalities and tumor types. PSO, with its efficient global search mechanism, automates this process to deliver consistently high segmentation accuracy. PSO iteratively explores the solution space, with each particle representing a configuration (e.g., filters, kernel size, learning rate), evaluated using the Dice Similarity Coefficient (DSC) in Eq.~\ref{eq_dsc}. 

\begin{equation} \label{eq_dsc}
\text{DSC} = \frac{2 |\mathbf{P} \cap \mathbf{T}|}{|\mathbf{P}| + |\mathbf{T}|}
\end{equation}

As illustrated in Figure~\ref{fig:unet_arch}, the PSO optimization process proceeds as follows:

\begin{enumerate}
    \item \textbf{Initialization}: Generate an initial swarm of particles, each representing a unique hyperparameter configuration (filters, kernel size, learning rate).
    \item \textbf{Evaluation}: Train the U-Net model corresponding to each particle’s configuration and evaluate segmentation accuracy using DSC on the validation set.
    \item \textbf{Update personal and global best}: For each particle, update the personal best (best performance achieved by that particle) and global best (best performance across all particles).
    \item \textbf{Update velocity and position}: PSO updates particle velocities and positions, guided by cognitive and social factors.
\end{enumerate}

The PSO process in Algorithm~\ref{alg_pso} includes initialization, fitness evaluation, and updates for velocity (Eq.\ref{eq_06}) and position (Eq.\ref{eq_07}), guiding the swarm toward the optimal configuration by balancing exploration and exploitation~\cite{Saifullah2024_mdpi}.

\begin{equation} \label{eq_06}
\mathbf{v}_i(t+1) = w \cdot \mathbf{v}_i(t) + c_1 \cdot r_1 \cdot (\mathbf{p}_{\text{best}, i} - \mathbf{x}_i(t)) + c_2 \cdot r_2 \cdot (\mathbf{g}_{\text{best}} - \mathbf{x}_i(t))
\end{equation}

\begin{equation} \label{eq_07}
\mathbf{x}_i(t+1) = \mathbf{x}_i(t) + \mathbf{v}_i(t+1)
\end{equation}

Here, \( v_i(t) \) and \( x_i(t) \) represent the velocity and position (hyperparameters) of particle \( i \) at iteration \( t \); \( w \), \( c_1 \), and \( c_2 \) are inertia weight, cognitive, and social coefficients, respectively; and \( r_1 \), \( r_2 \) are random values in \([0,1]\).

\begin{algorithm} [ht]
\caption{Particle Swarm Optimization (PSO)}
\label{alg_pso}
\begin{algorithmic}[1]
\STATE Initialize particles: $\mathbf{x}_i^{(0)}, \mathbf{v}_i^{(0)} \quad \forall i = 1, \dots, N$
\STATE Initialize personal best: $\mathbf{p}_i^{best} = \mathbf{x}_i^{(0)} \quad \forall i$
\STATE Initialize global best: $\mathbf{g}^{best} = \arg \min (\{ f(\mathbf{p}_i^{best}) \}_{i=1}^N)$
\STATE Set parameters: $w, c_1, c_2$ AND maximum iterations: $T$

\FOR{$t = 1$ \textbf{to} $T$}
    \FOR{$i = 1$ \textbf{to} $N$}
        \STATE Update velocity: using Eq.~\ref{eq_06}
        \STATE Update position: using Eq.~\ref{eq_07}
        \STATE Evaluate fitness: $f(\mathbf{x}_i^{(t)})$
        \IF{$f(\mathbf{x}_i^{(t)}) < f(\mathbf{p}_i^{best})$}
            \STATE Update personal best: $\mathbf{p}_i^{best} = \mathbf{x}_i^{(t)}$
        \ENDIF
        \IF{$f(\mathbf{x}_i^{(t)}) < f(\mathbf{g}^{best})$}
            \STATE Update global best: $\mathbf{g}^{best} = \mathbf{x}_i^{(t)}$
        \ENDIF
    \ENDFOR
\ENDFOR

\STATE \textbf{Return} global best: $\mathbf{g}^{best}$, $f(\mathbf{g}^{best})$
\end{algorithmic}
\end{algorithm}

The iterative Particle Swarm Optimization (PSO) process effectively balances exploration and exploitation, enabling rapid convergence to optimal hyperparameters that improve U-Net’s segmentation accuracy and efficiency. The proposed PSO-UNet framework not only enhances segmentation accuracy but also introduces a systematic, lightweight hyperparameter optimization strategy that outperforms traditional tuning methods across diverse multimodal MRI datasets and tumor types. This integration ensures scalability and computational efficiency, making the framework adaptable to different dataset characteristics.

\subsection{Datasets and Preprocessing}
\label{3.2-data-prep}
\begin{figure}[ht]
  \centering
  \begin{subfigure}[b]{\linewidth}
    \centering
    \includegraphics[width=\linewidth]{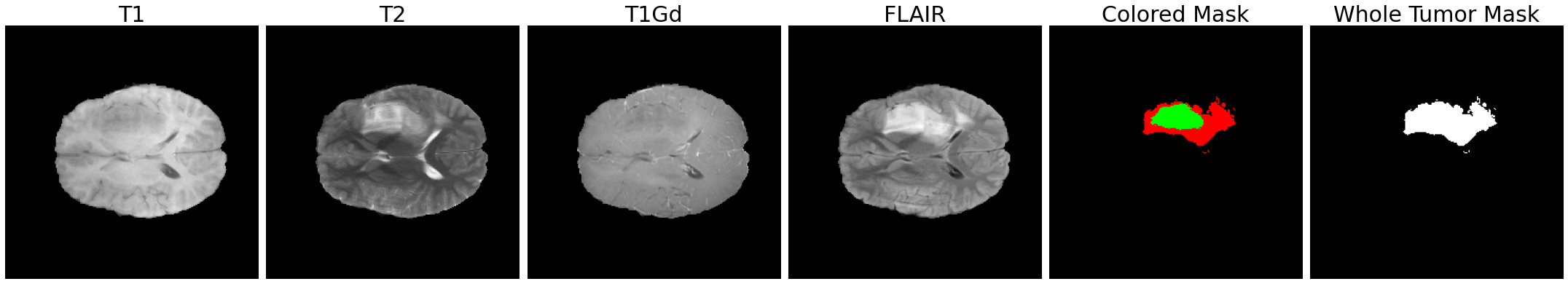}
    \caption{} 
    \label{fig:sample_brats}
  \end{subfigure}
  \begin{subfigure}[b]{\linewidth}
    \centering
    \includegraphics[width=\linewidth]{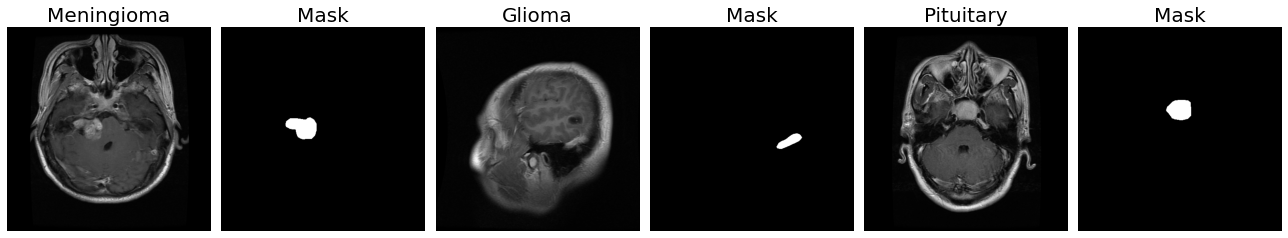}
    \caption{} 
    \label{fig:sample_figshare}
  \end{subfigure}
  \caption{MRI images from (a) BraTS 2021 and (b) Figshare with their masks.}
  \label{fig:dts_1}
\end{figure}

The PSO-UNet framework was evaluated on BraTS 2021~\cite{Baid2023} and the Figshare~\cite{Cheng2017} brain tumor dataset, selected for their multimodal and multi-class segmentation challenges. The BraTS 2021 contains 1,251 MRI samples across four modalities (T1, T2, T1Gd, FLAIR), each with masks tumor annotations (Fig.~\ref{fig:sample_brats}), focusing on segmenting the Whole Tumor (WT) region. The Figshare dataset, contributed by Cheng et al., includes 3,064 MRI scans across three tumor types (Meningioma, Glioma, Pituitary) with the masks (Fig.~\ref{fig:sample_figshare}), introducing complexity through inter-class variability, diverse tumor geometries, and intensity differences.

To ensure consistent performance across datasets, we applied the following preprocessing steps:

\begin{enumerate}
    \item \textbf{Resizing}: Resized MRI images to $256 \times 256$ for stable modeling~\cite{Saifullah2025}.
    \item \textbf{Conversion}: Converted MRI scans to uint8 format to reduce memory usage.
    \item \textbf{Channel Duplication}: Duplicated grayscale images across three channels.
    \item \textbf{Mask Normalization}: Normalized ground truth masks to binary values ($0$ for background, $1$ for tumor).
\end{enumerate}

These preprocessing steps mitigated dataset variability and ensured consistent model performance. Standardized inputs enabled PSO-UNet to generalize effectively across imaging modalities and tumor types for accurate segmentation.

\subsection{Implementation and Evaluation Metrics}
\label{3.3-implementation-evaluation}
The PSO-UNet framework was developed using TensorFlow 2.10 and utilizes Particle Swarm Optimization (PSO) to optimize the model's hyperparameters for better segmentation accuracy and efficiency. The framework was trained on an 8x NVIDIA A100-SXM4-40GB GPU for 50 epochs, with a batch size of 16 per PSO iteration. The training process used a combined loss function, integrating Dice Loss and Cross-Entropy Loss with weighting factors $\alpha$ and $\beta$ (Eq.~\ref{eq_loss}), to balance regional overlap and pixel-wise accuracy. The dataset was split 80\% for training and 20\% for validation, ensuring adequate training and evaluation sets.

\begin{equation} \label{eq_loss}
\mathcal{L}_{\text{total}} = \alpha \cdot \mathcal{L}_{\text{Dice}} + \beta \cdot \mathcal{L}_{\text{CrossEntropy}}
\end{equation}

PSO optimized three key hyperparameters: the number of filters ([8, 16, 32, 64]), kernel size ([3, 4, 5]), and learning rate ([0.0001, 0.01]). Each particle in the swarm represented a unique configuration, with fitness evaluated using segmentation metrics like DSC (Eq.~\ref{eq_dsc}). Additionally, Intersection over Union (IoU, Eq.~\ref{eq_iou}) was used to assess the ratio of the intersection to the union of predicted ($\mathbf{P}$) and ground truth ($\mathbf{T}$) regions, where a higher IoU indicates better segmentation accuracy. The details of the configuration of PSO optimization process are shown in Table~\ref{tbl_conf}.

\begin{equation} \label{eq_iou}
\text{IoU} = \frac{|\mathbf{P} \cap \mathbf{T}|}{|\mathbf{P} \cup \mathbf{T}|}
\end{equation}

\begin{table}[h]
\caption{Configuration of PSO for Hyperparameter Tuning}
\label{tbl_conf}
\centering
\begin{tabular}{lll}
\hline
\textbf{Category} & \textbf{Parameter} & \textbf{Value/Description} \\ \hline
            & filter\_values & [8, 16, 32, 64] \\
            && (Possible filter values) \\ 
            & lb & [0, 3, 0.0001] \\
            &&(Lower bounds for filter index, \\
            && kernel size, and learning rate) \\ 
            & ub & [3, 5, 0.01] \\
PSO         && (Upper bounds for filter index, \\
Confi-      && kernel size, and learning rate) \\ 
guration    & options & {c1: 0.5, c2: 0.3, w: 0.9} \\
            &&(PSO configuration options for \\
            &&cognitive and social factors, \\
            && and inertia weight) \\ 
            & n\_particles & 10 \\
            &&(Number of particles for PSO \\
            && optimization process) \\ \hline
PSO         & iters & 10 \\
Optimi-     && (Number of iterations for \\
zation      && PSO search) \\ \hline
            & obj\_func & Wrapper for evaluating the  \\
Objective   && model with varying hyper- \\
Function    && parameters (filters, kernel \\ 
            && size, learning rate) \\ \hline
\end{tabular}
\end{table}

Pixel-wise classification performance was evaluated using Accuracy (Eq.~\ref{eq_acc}). True positives (TP), true negatives (TN), false positives (FP), and false negatives (FN) are used to compute the proportion of correctly classified pixels.

\begin{equation} \label{eq_acc}
\text{Accuracy} = \frac{\text{TP} + \text{TN}}{\text{TP} + \text{TN} + \text{FP} + \text{FN}}
\end{equation}

Boundary precision was evaluated using advanced metrics such as the Hausdorff Distance (HD) and Average Symmetric Surface Distance (ASSD). HD (Eq.~\ref{eq_hd}) measures the maximum boundary deviation between predicted and ground truth masks, making it sensitive to outliers, while ASSD (Eq.~\ref{eq_assd}) computes the average boundary deviation, offering a balanced measure by averaging the Euclidean distances between corresponding boundary points.

\begin{equation} \label{eq_hd}
\text{HD} = \max \big( \sup_{p \in \mathbf{P}} \inf_{t \in \mathbf{T}} d(p, t), \sup_{t \in \mathbf{T}} \inf_{p \in \mathbf{P}} d(t, p) \big)
\end{equation}

\begin{equation} \label{eq_assd}
\text{ASSD} = \frac{1}{|\mathbf{P}| + |\mathbf{T}|} \Big( \sum_{p \in \mathbf{P}} \inf_{t \in \mathbf{T}} d(p, t) + \sum_{t \in \mathbf{T}} \inf_{p \in \mathbf{P}} d(t, p) \Big)
\end{equation}

Precision, Recall, and F1 Score were used to assess segmentation performance. Precision (Eq.~\ref{eq_precision}) measures true positives among positive predictions, Recall (Eq.~\ref{eq_recall}) quantifies sensitivity to actual positives, and the F1 Score (Eq.~\ref{eq_f1}) is the harmonic mean of Precision and Recall.

\begin{equation} \label{eq_precision}
\text{Precision} = \frac{\text{TP}}{\text{TP} + \text{FP}}
\end{equation}

\begin{equation} \label{eq_recall}
\text{Recall} = \frac{\text{TP}}{\text{TP} + \text{FN}}
\end{equation}

\begin{equation} \label{eq_f1}
\text{F1 Score} = 2 \cdot \frac{\text{Precision} \cdot \text{Recall}}{\text{Precision} + \text{Recall}}
\end{equation}

\section{Results and Discussion}
\label{results-and-discussion}
This section presents PSO-UNet's results for medical image segmentation, including optimized U-Net parameters via PSO, evaluation of multimodal and multi-class MRI datasets, and comparison with state-of-the-art methods, highlighting high metrics and computational efficiency. The experimental results validate the effectiveness of integrating Particle Swarm Optimization (PSO) into U-Net, as illustrated in Figure~\ref{fig:unet_arch}, where PSO dynamically tunes filters (\( f \)) and kernel size (\( k \)) within the encoder and decoder blocks.

\subsection{PSO: Critical Analysis of UNet Architecture Parameters}
\label{4.1-pso-analysis}

The Particle Swarm Optimization (PSO) algorithm plays a crucial role in optimizing key architectural parameters of the U-Net for medical image segmentation. By balancing exploration and exploitation, the PSO efficiently traverses the hyperparameter space to identify configurations that maximize segmentation accuracy and model generalization.

As detailed in Section~\ref{3.3-implementation-evaluation}, the swarm systematically explored hyperparameter configurations, including the number of filters \( f \in \{8, 16, 32, 64\} \), kernel sizes \( k \in \{3, 4, 5\} \), and learning rates ranging from 0.0001 to 0.01. This dynamic exploration enabled PSO to refine model configurations effectively during training.

\begin{figure}
\centering
    \begin{tabular}{c}
      \includegraphics[width=.5\textwidth]{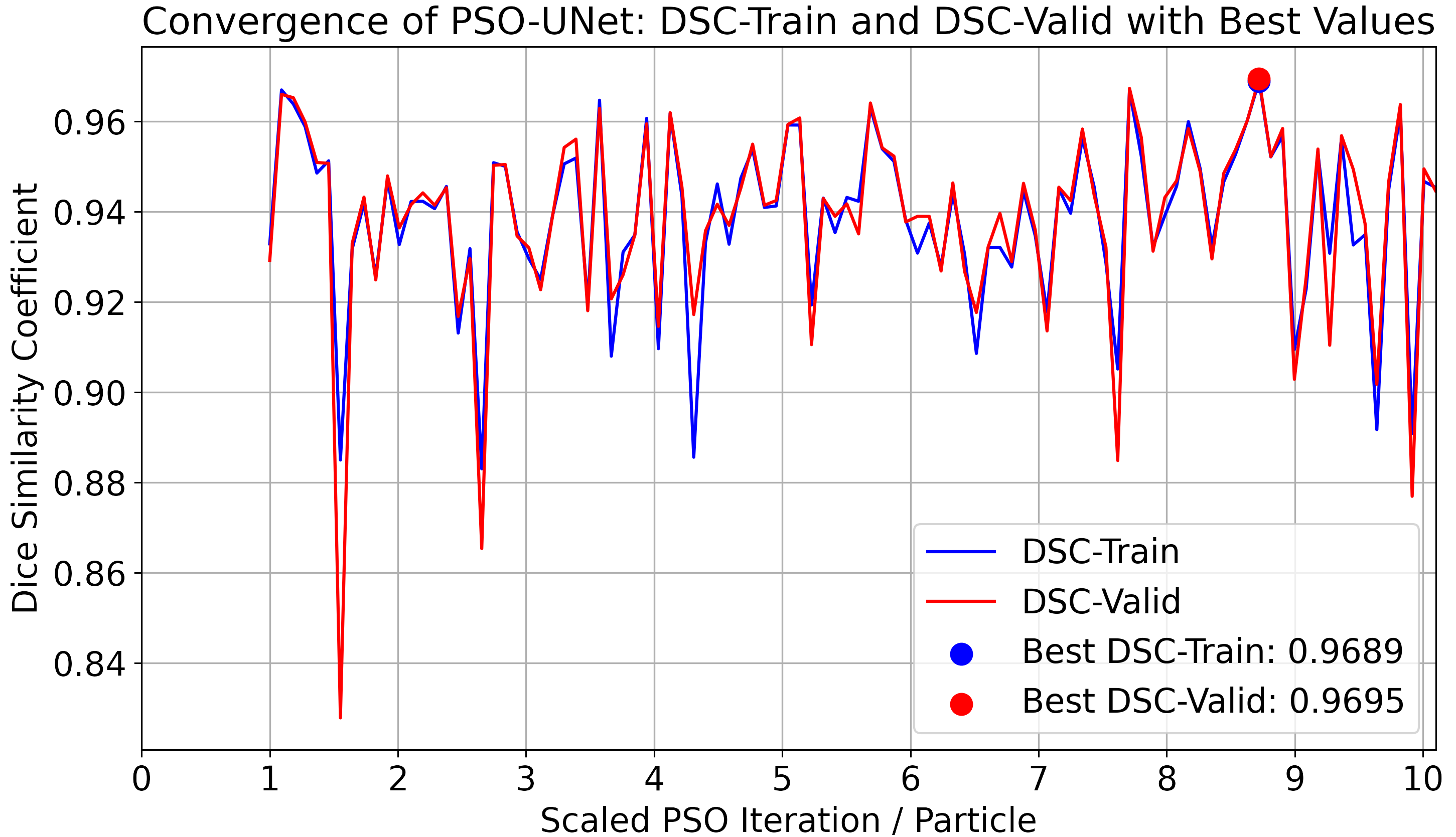}       \\
      (a) \\
      \includegraphics[width=.5\textwidth]{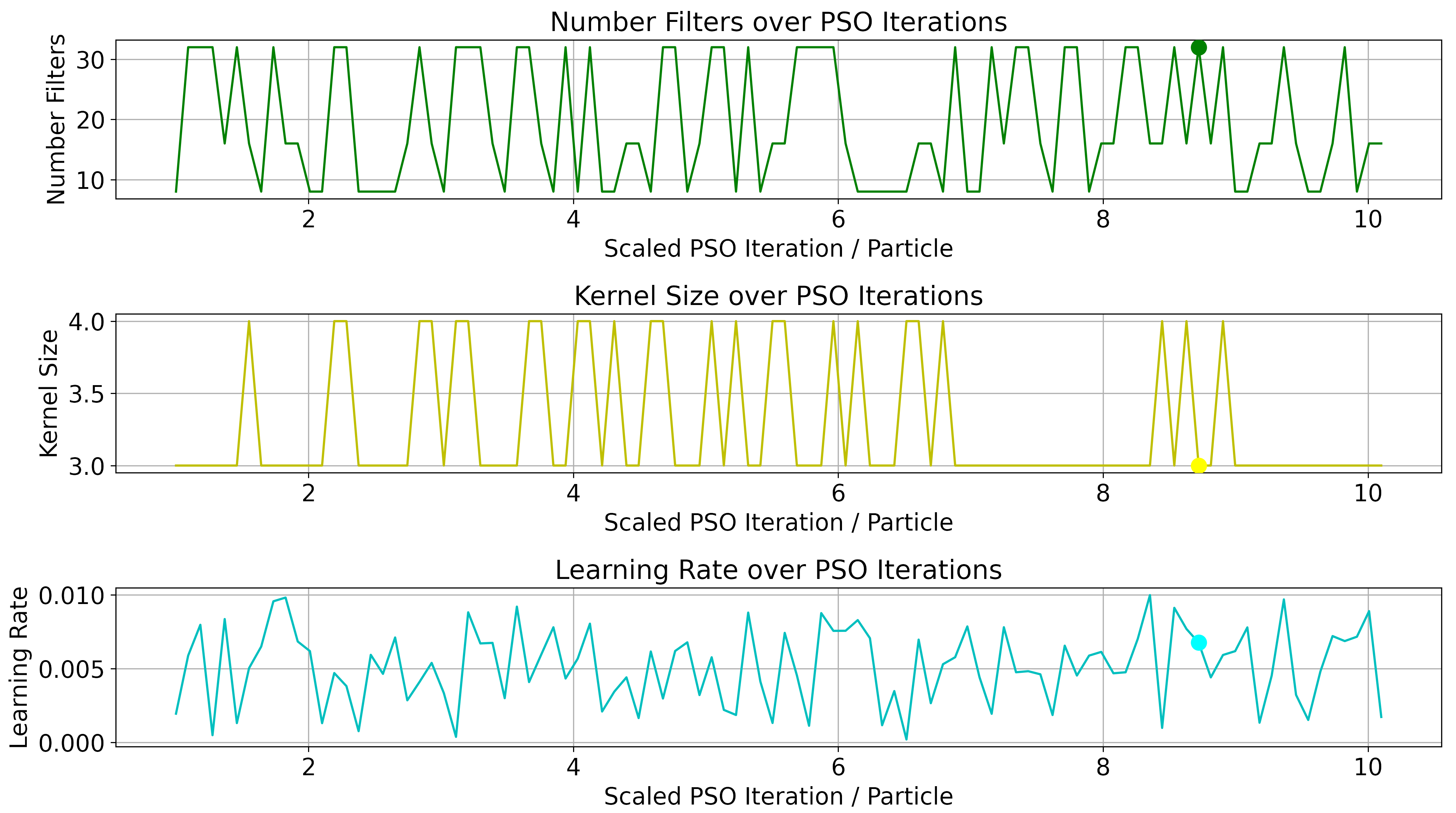}\\
      (b)\\
      \includegraphics[width=.4\textwidth]{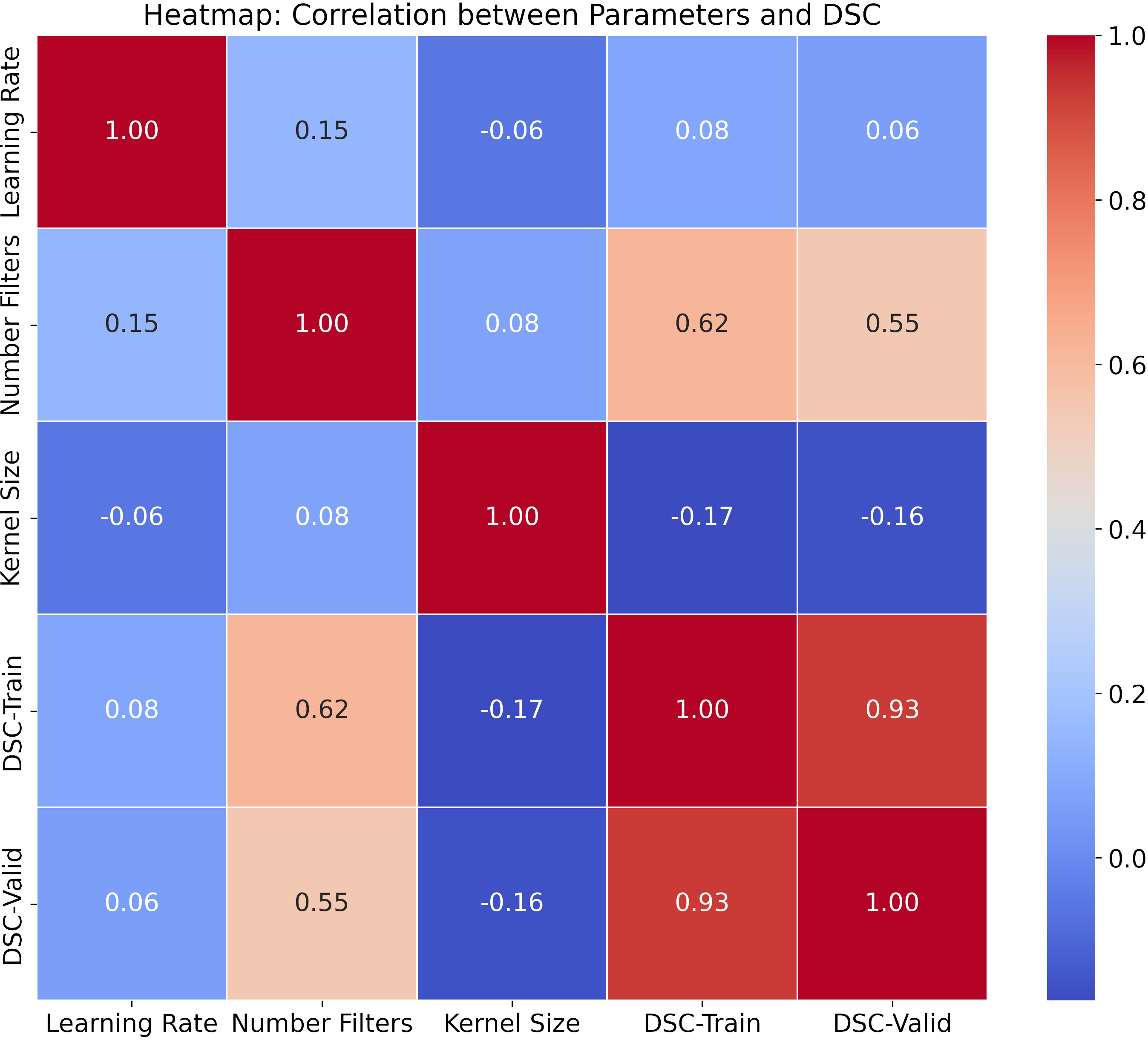}\\
      (c)\\
          
    \end{tabular}
\caption{(a) PSO-UNet convergence: DSC (train in blue, validation in red); (b) Evolution of architecture parameters (filters, kernel size, learning rate) across 10 particles and generations; (c) Correlation heatmap of evaluation metrics.}
\label{fig:convergence_param}
\end{figure}

Figure~\ref{fig:convergence_param}a shows the convergence of the Dice Similarity Coefficients (DSC) for training and validation, peaking at 0.9689 and 0.9695 in generation 4. Early fluctuations in DSC indicate the exploration phase of PSO, where diverse combinations of parameters are evaluated. By generation 4, the algorithm transitions to the exploitation phase, focusing on fine-tuning the best configurations for improved stability. This balance between exploration and exploitation helps the algorithm avoid local optima and ensures robust performance across training and validation datasets.

Figure~\ref{fig:convergence_param}b highlights the evolution of key parameters: The convergence to 32 filters emphasizes their critical role in feature extraction, with larger filters enabling better capture of complex patterns. The kernel size of 3 is favored for capturing fine spatial details, making it effective for precise segmentation. The learning rate converges to 0.00675, balancing training efficiency and stability, ensuring that the model converges smoothly without oscillations. These findings demonstrate how PSO systematically refines each parameter to achieve optimal performance.

Figure~\ref{fig:convergence_param}c shows a strong positive correlation (0.93) between DSC-Train and DSC-Valid, confirming the model's generalization ability. The learning rate is the most influential parameter, followed by the filter count, indicating their significant role in achieving high performance. The kernel size, although important, plays a secondary role, as evidenced by its weaker correlation with the DSC metrics. These results validate PSO’s effectiveness in optimizing the U-Net architecture for high performance and strong generalization to unseen data.

\subsection{Evaluation of the Proposed PSO-UNet Framework Across Modalities and Tumor Classes}
\label{4.2-pso-unet-evaluation}
\begin{figure}
\centering
\begin{tabular}{c}
     \includegraphics[width=.4\textheight]{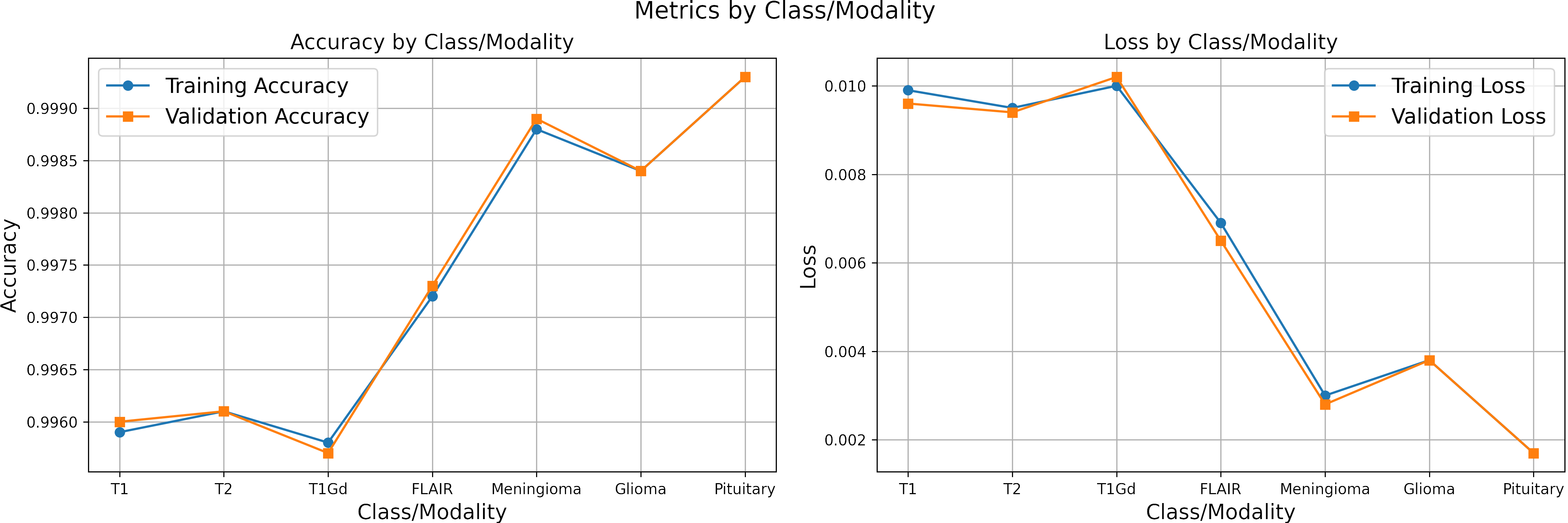} \\ \includegraphics[width=.4\textheight]{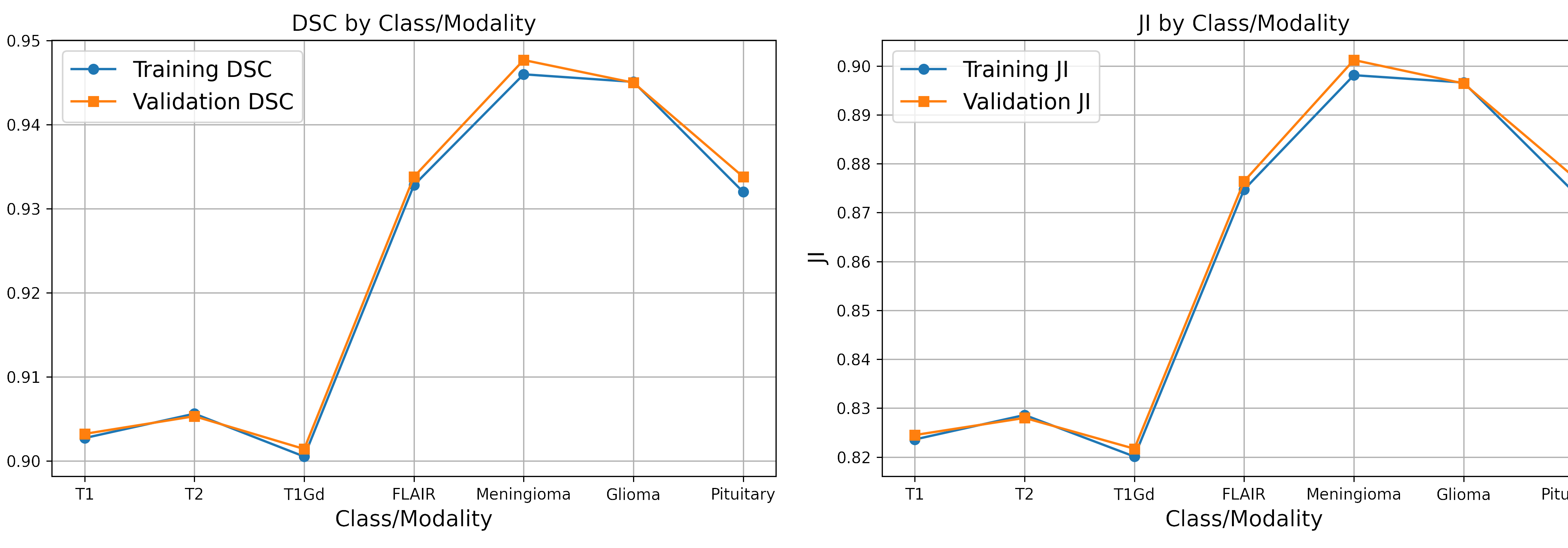}\\ 
\end{tabular} 
\caption{Evaluation metrics (Accuracy, Loss, DSC, and IoU) for the PSO-UNet framework across BraTS 2019 modalities (T1, T2, T1Gd, FLAIR) and Figshare tumor classes (Meningioma, Glioma, Pituitary).} \label{fig:evaluation} \end{figure}

The PSO-UNet framework was evaluated using optimal hyperparameters (Number Filters = 32, Kernel Size = 3, Learning Rate = 0.006756673) applied to the BraTS 2019 dataset (T1, T2, T1Gd, FLAIR modalities) and the Figshare dataset (Meningioma, Glioma, Pituitary tumor classes). Performance was assessed using Accuracy, Loss, Dice Similarity Coefficient (DSC), and Intersection over Union (IoU), with results shown in Figure~\ref{fig:evaluation}. The performance trends for these metrics demonstrate strong consistency between training and validation, indicating robust generalization. For instance, the FLAIR validation accuracy is 0.9973, close to the training accuracy (0.9972), and Meningioma's validation IoU is 0.9012, slightly higher than the training IoU (0.8981). These results confirm that the PSO-UNet framework avoids overfitting and maintains reliable performance on unseen data.

The high DSC and IoU values across most modalities and tumor classes demonstrate the PSO-UNet framework's effectiveness in medical image segmentation. FLAIR modality achieved the highest validation DSC of 0.9338 and IoU of 0.8764, benefiting from its ability to suppress cerebrospinal fluid signals and delineate tumor boundaries. T1 modality performed slightly lower, with a validation DSC of 0.9032 and IoU of 0.8245, likely due to its limited ability to highlight tumor-specific features. T2 and T1Gd modalities also demonstrated good performance, with DSCs of 0.9053 and 0.9014, respectively, and IoUs of 0.8280 and 0.8217, reflecting their ability to enhance tumor region visibility, particularly for vascularized regions in the case of T1Gd. While T1Gd is valuable for identifying vascularized tumor regions, suboptimal contrast enhancement in some cases can affect its performance.

The model performed best on Meningioma, with the highest validation DSC (0.9477) and IoU (0.9012), due to well-defined tumor boundaries. Glioma and Pituitary also showed strong results, with DSC values of 0.9450 and 0.9338, and IoUs of 0.8964 and 0.8774. These findings demonstrate the model’s ability to handle both localized and diffuse tumor types. Consistent training and validation metrics, such as a 0.1 p.p. (percentage points) DSC difference for FLAIR and a slight increase in Meningioma validation IoU, show PSO-UNet’s generalization capability. Low validation losses (e.g., 0.0017 for Pituitary) further confirm the model’s accuracy and training stability. These results confirm that PSO-UNet achieves state-of-the-art segmentation performance with strong generalization across tumor types and imaging modalities, ensuring its reliability for clinical applications.

\subsection{Segmentation Results of Predicted Models}
\label{4.3-segmentation-results}
\begin{figure}
\centering
\begin{tabular}{cccc}
\begin{tabular}[c]{@{}c@{}}\includegraphics[width=.1\textwidth]{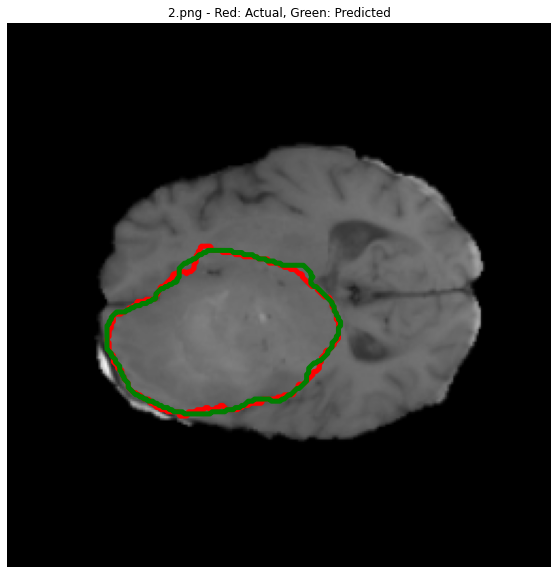}\end{tabular} & 
\begin{tabular}[c]{@{}c@{}}\includegraphics[width=.1\textwidth]{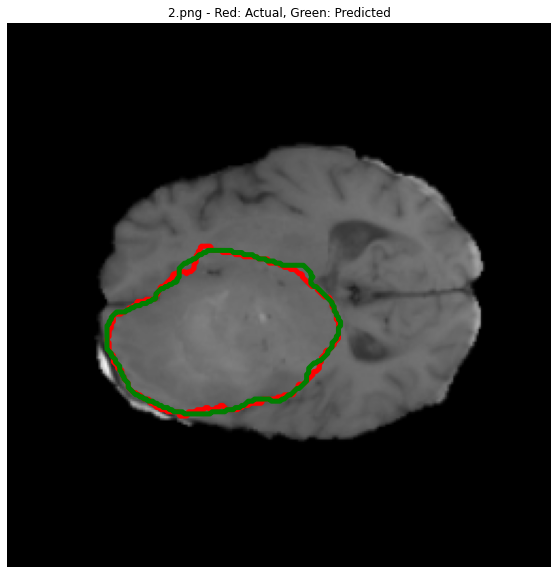}\end{tabular} & 
\begin{tabular}[c]{@{}c@{}}\includegraphics[width=.1\textwidth]{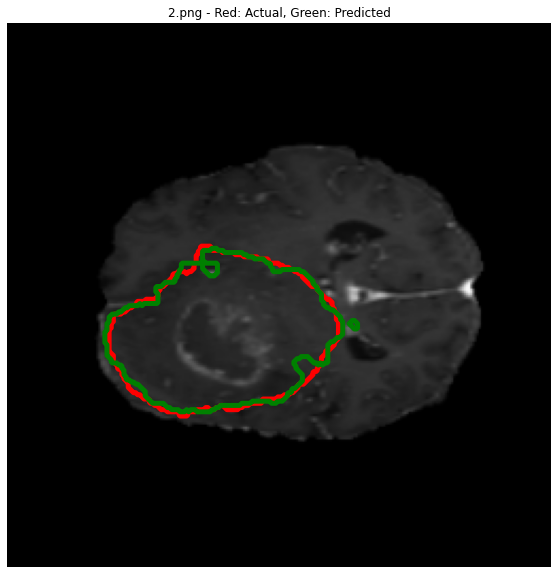}\end{tabular} &
\begin{tabular}[c]{@{}c@{}}\includegraphics[width=.1\textwidth]{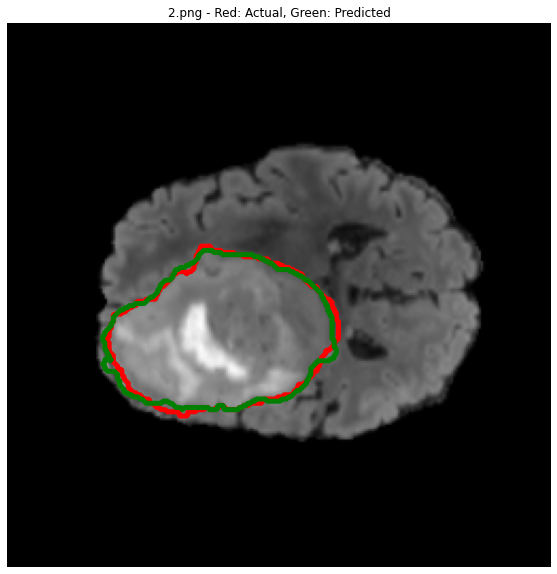}\end{tabular}\\ 

\begin{tabular}[c]{@{}c@{}}\scriptsize T1 \\ \scriptsize DSC: 0.9743 \\ \scriptsize IoU: 0.9499 \\ \scriptsize HD: 4.1231 \\ \scriptsize ASSD: 0.0372\end{tabular} 
& \begin{tabular}[c]{@{}c@{}}\scriptsize T2 \\ \scriptsize DSC: 0.9743 \\ \scriptsize IoU: 0.9499 \\ \scriptsize HD: 4.1231 \\ \scriptsize ASSD: 0.0372 \\\end{tabular} 
& \begin{tabular}[c]{@{}c@{}}\scriptsize T1Gd \\ \scriptsize DSC: 0.9556 \\ \scriptsize IoU: 0.9150 \\ \scriptsize HD: 9.0000 \\ \scriptsize ASSD: 0.0761 \end{tabular} 
& \begin{tabular}[c]{@{}c@{}} \scriptsize FLAIR \\ \scriptsize DSC: 0.9651\\ \scriptsize IoU: 0.9325\\ \scriptsize HD: 5.0990\\ \scriptsize ASSD: 0.0571 \end{tabular} \\
\end{tabular}

\begin{tabular}{ccc}
\begin{tabular}[c]{@{}c@{}}\includegraphics[width=.1\textwidth]{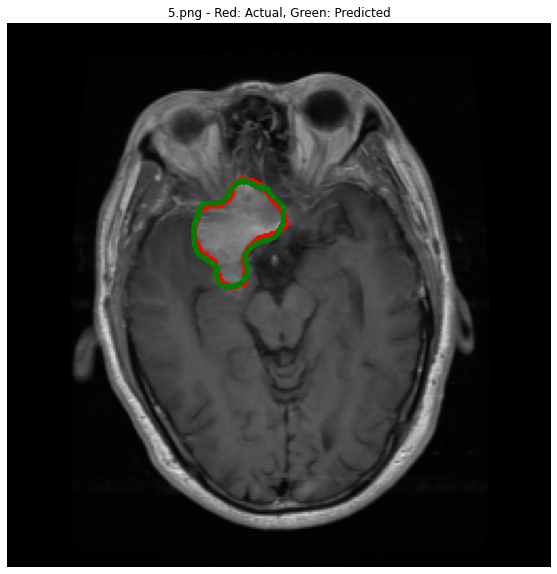}\end{tabular} & 
\begin{tabular}[c]{@{}c@{}}\includegraphics[width=.1\textwidth]{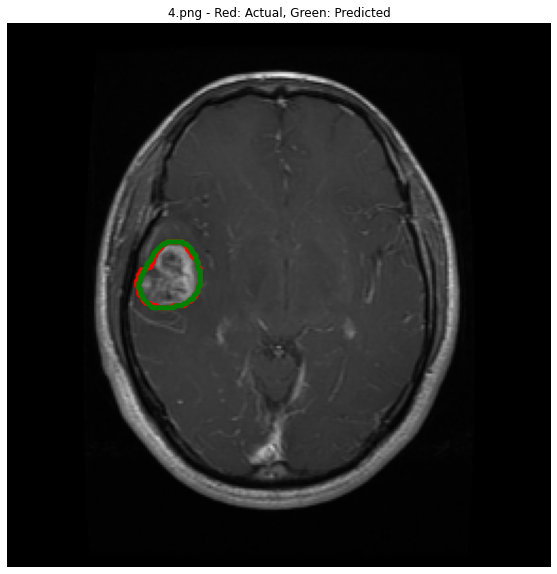}\end{tabular} & 
\begin{tabular}[c]{@{}c@{}}\includegraphics[width=.1\textwidth]{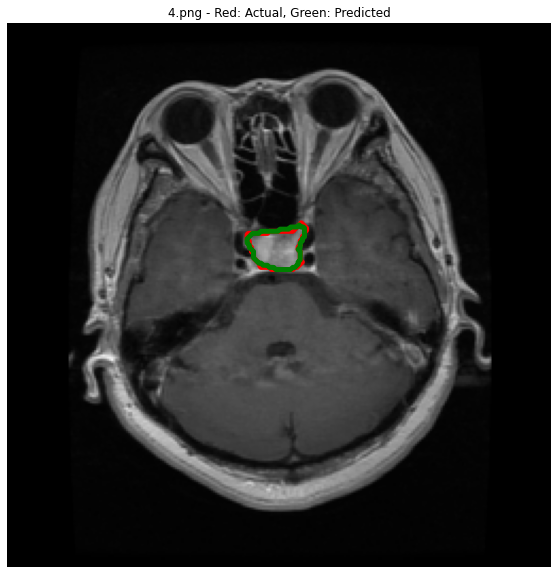}\end{tabular} \\

\begin{tabular}[c]{@{}c@{}}\scriptsize Meningioma \\ \scriptsize DSC: 0.9450 \\ \scriptsize IoU: 0.8958 \\ \scriptsize HD: 2.0000 \\ \scriptsize ASSD: 0.0616 \end{tabular} 
& \begin{tabular}[c]{@{}c@{}}\scriptsize Glioma \\ \scriptsize DSC: 0.9643 \\ \scriptsize IoU: 0.9310 \\ \scriptsize HD: 2.0000 \\ \scriptsize ASSD: 0.0383\end{tabular} 
& \begin{tabular}[c]{@{}c@{}}\scriptsize Pituitary \\ \scriptsize DSC: 0.9494 \\ \scriptsize IoU: 0.9038 \\ \scriptsize HD: 2.0000 \\ \scriptsize ASSD: 0.0543 \end{tabular} 
\\
\end{tabular}

\caption{Performance Metrics for Predicted vs. Mask Overlaps} \label{figsample3}
\end{figure}

The PSO-UNet framework was evaluated by comparing predicted segmentation masks with ground truth across various modalities (T1, T2, T1Gd, FLAIR) and tumor classes (Meningioma, Glioma, Pituitary). Key metrics such as Dice Similarity Coefficient (DSC), Intersection over Union (IoU), Hausdorff Distance (HD), and Average Symmetric Surface Distance (ASSD) were used to assess segmentation quality. Figure~\ref{figsample3} visualizes segmentation performance, while Table~\ref{tab:metrics_evaluation} summarizes Accuracy, Precision, Recall, and F1 Score.

Among the modalities, T1 and T2 achieved the highest DSC (0.9743) and IoU (0.9499), along with the lowest HD (4.1231) and ASSD (0.0372), indicating robust performance in accurately capturing tumor regions. FLAIR modality showed good performance with DSC (0.9651) and IoU (0.9325), although its HD (5.0990) and ASSD (0.0571) were slightly higher, suggesting minor boundary discrepancies. T1Gd exhibited comparatively lower performance with DSC (0.9556) and IoU (0.9150), likely due to imaging artifacts and contrast variability, which hindered identification of tumor boundary.

\begin{table}[h]
\caption{Segmentation Performance Metrics for Different Modalities and Tumor Classes.}
\label{tab:metrics_evaluation}
\centering
\begin{tabular}{lcccc}
\hline
\textbf{Modality/Class} & \textbf{Accuracy} & \textbf{Precision} & \textbf{Recall} & \textbf{F1 Score} \\
\hline
T1         & 0.9960 & 0.9311 & 0.9415 & 0.9362 \\
T2         & 0.9961 & 0.9528 & 0.9205 & 0.9364 \\
T1Gd       & 0.9945 & 0.9265 & 0.8944 & 0.9102 \\
FLAIR      & 0.9968 & 0.9547 & 0.9405 & 0.9475 \\
\hline
Meningioma & 0.9989 & 0.9675 & 0.9696 & 0.9685 \\
Glioma     & 0.9984 & 0.9738 & 0.9519 & 0.9627 \\
Pituitary  & 0.9993 & 0.9596 & 0.9593 & 0.9594 \\
\hline
\end{tabular}
\end{table}

\begin{figure}[ht]
\centering
\includegraphics[width=.45\textwidth]{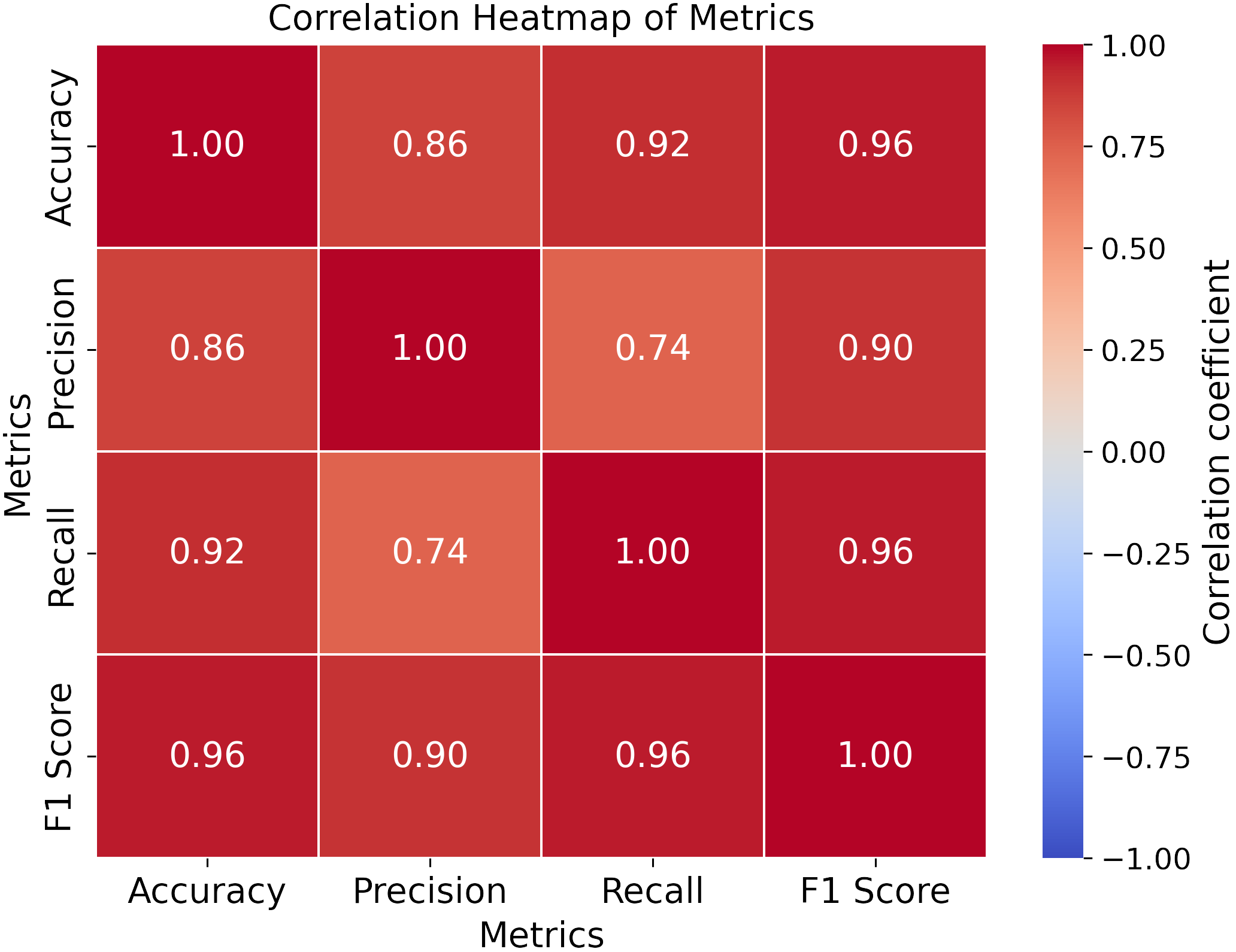}
\caption{Correlation heatmap of evaluation metrics of Table~\ref{tab:metrics_evaluation}.}
\label{fig:heatmap}
\end{figure}

Regarding tumor classes, Glioma achieved the highest DSC (0.9643) and IoU (0.9310), demonstrating the model's strength in handling diffuse tumor boundaries. Meningioma showcased the model’s capability in segmenting well-defined, localized tumors with DSC of 0.9450 and IoU of 0.8958. Despite its smaller and more challenging nature, the Pituitary tumor achieved DSC of 0.9494 and IoU of 0.9038, indicating the framework’s reliability even in more compact and complex segmentation tasks.

Table~\ref{tab:metrics_evaluation} further highlights performance with metrics such as Accuracy, Precision, Recall, and F1 Score. FLAIR achieved the highest accuracy (0.9968) and F1 Score (0.9475), while Glioma had the highest Precision (0.9738) and F1 Score (0.9627), underscoring its effectiveness in minimizing false positives. Figure~\ref{fig:heatmap} shows a strong correlation between Accuracy and F1 Score (0.96) and Recall and F1 Score (0.96) but a weaker correlation between Precision and Recall (0.74), suggesting room for improvement in handling boundary alignment, especially for T1Gd. These results confirm the framework's robustness and potential for clinical applications in medical image segmentation.

\subsection{Evaluation of the Proposed Method Compared with State-of-the-Art Techniques}
\label{4.4-sota}
The PSO-UNet framework was evaluated on the BraTS 2021 dataset (Whole Tumor region) and the Figshare dataset (Meningioma, Glioma, Pituitary tumor classes), using Dice Similarity Coefficient (DSC) and Intersection over Union (IoU). The results, summarized in Table~\ref{tab:comparison}, provide a comparative analysis of its performance against state-of-the-art segmentation methods.

\begin{table}[ht]
\caption{Quantitative Comparison of PSO-UNet with Other Methods}
\label{tab:comparison}
\centering
\begin{tabular}{lcc}
\hline
\textbf{Method} & \textbf{DSC} & \textbf{IoU} \\
\hline
\multicolumn{3}{c}{\textbf{BraTS 2021 Dataset}} \\
\hline
\textbf{Proposed Method} & \textbf{0.9578} & \textbf{0.9194}  \\
ViT-self-attention~\cite{Ghazouani2023} & 0.9176 & - \\
UNet~\cite{Hernandez-Gutierrez2024} & 0.8600 & 0.7807 \\
MS-SegNet~\cite{Sachdeva2024} & 0.9200 & - \\
ResU-Net~\cite{Bouchet2023} & 0.8841 & - \\
ViT-24~\cite{Mojtahedi2023} & 0.8048 & - \\
SPPNet-2~\cite{Vijay2023} & 0.9040 & - \\
DenseUNet+~\cite{Cetiner2023} & 0.9500 & 0.8800 \\
U-Net\_ASPP\_EVO~\cite{Yousef2023} & 0.9251 & - \\
UNCE-NODE~\cite{Sadique2023} & 0.8949 & - \\
2-Cascaded U-Net~\cite{Jiang2020} & 0.8370 & - \\
ViT-32~\cite{Mojtahedi2023} & 0.8045 & - \\
Att-ResU-Net~\cite{Bouchet2023} & 0.8602 &- \\ 
SPPNet-1~\cite{Vijay2023} & 0.8999 & -\\

\hline
\multicolumn{3}{c}{\textbf{Figshare Dataset}} \\
\hline
\textbf{Proposed Method} & \textbf{0.9523} & \textbf{0.9097} \\
UNet-T-PSO~\cite{Saifullah2024} & 0.9312 & 0.8722 \\
ResUnet-TL~\cite{Murmu2023} & 0.9194 & 0.9495 \\
DLV3+ResNet18~\cite{Saifullah2024_ijece} & 0.9124 & 0.9340 \\
MST-based~\cite{Mayala2022} & 0.8469 & 0.7443 \\
ResUNet101~\cite{Ingle2022} & 0.8369 & 0.8500 \\
Modified ResNet (CRF-TL)~\cite{Chatterjee2024} & 0.8300 & 0.8400 \\
Ensemble UNet-ResNet~\cite{Roshan2024} & 0.8731 & 0.7902 \\
EAV-UNet~\cite{Cheng2023} & 0.7680 & 0.8420 \\
CD-SL~\cite{Sobhaninia2020} & 0.8003 & 0.9074 \\
KFCM-CNN~\cite{Rao2019} & 0.8884 & 0.8204 \\
Mask RCNN~\cite{Masood2021} & 0.9500 & - \\
CCN-PR-Seg-net~\cite{Tripathi2021} & 0.9330 & 0.8820 \\
Pipeline U-Net~\cite{KumarBhatt2023} & - & 0.8312 \\
\hline
\end{tabular}
\end{table}

On the BraTS 2021 dataset, PSO-UNet achieved DSC of 0.9578 and IoU of 0.9194, showing 0.78 p.p. (percentage points) improvement in DSC and 4.45 p.p. in IoU compared to DenseUNet+ (DSC: 0.9500, IoU: 0.8800). Models like UNet (DSC: 0.8600, IoU: 0.7807) and ViT-self-attention (DSC: 0.9176) demonstrated lower performance due to static configurations or limited optimization strategies. Other methods, such as MS-SegNet and SPPNet-2, also achieved lower DSC values of 0.9200 and 0.9040, respectively, highlighting the advantage of PSO-UNet’s dynamic hyperparameter tuning in achieving superior segmentation accuracy.

In contrast, on the Figshare dataset, PSO-UNet achieved a DSC of 0.9523 and IoU of 0.9097, which represents 2.11 p.p. higher DSC and 3.74 p.p. higher IoU compared to UNet-T-PSO (DSC: 0.9312, IoU: 0.8722). Other methods, such as ResUnet-TL (DSC: 0.9194) and DLV3+ ResNet18 (DSC: 0.9124), also had lower IoU values, indicating less precise boundary segmentation. Traditional approaches like MST-based and ResUNet101, with DSC values of 0.8469 and 0.8369, showed significant limitations in handling complex tumor morphologies.

PSO-UNet's performance reflects its dynamic hyperparameter optimization, which adapts effectively to different tumor types and imaging modalities. The framework achieves precise boundary alignment by capturing fine-grained tumor features and minimizing false positives and negatives. Potential enhancements, such as incorporating attention mechanisms or hybrid optimization approaches, could improve its segmentation performance and scalability, particularly for challenging cases like T1Gd.

\subsection{Computational Efficiency and Evaluation of the Proposed Method}
\label{4.5-computational-evaluation}
The PSO-UNet framework achieves a remarkable balance between segmentation quality and computational efficiency, as summarized in Table~\ref{tab:efficiency_comparison}. It achieves a DSC of 0.9523 and IoU of 0.9097, with a parameter count of 7,771,873 and an execution time of 906 seconds. In comparison, UNet-T-PSO~\cite{Saifullah2024} and M-Unet~\cite{Saifullah2024_gecco}, which had 31,047,105 parameters, took 1711 and 1760 seconds to execute, respectively, resulting in a 46.9\% and 48.5\% reduction in execution time. These models achieve lower DSC values (0.9255 and 0.9220) and IoU scores (0.8628 and 0.8569), highlighting PSO-UNet's superior efficiency and segmentation quality.

\begin{table}[ht]
\caption{Computational Efficiency and Evaluation Metrics of PSO-UNet Compared with Other Methods}
\label{tab:efficiency_comparison}
\centering
\begin{tabular}{lrrcc}
\hline
\textbf{Method} & \textbf{Parameters} & \textbf{Times (s)} & \textbf{DSC} & \textbf{IoU} \\
\hline
\textbf{Proposed } & \textbf{7,771,873} & \textbf{906} & \textbf{0.9523} & \textbf{0.9097} \\
UNet-T-PSO~\cite{Saifullah2024} & 31,047,105 & 1711 & 0.9255 & 0.8628 \\
M-Unet~\cite{Saifullah2024_gecco} & 31,047,105 & 1760 & 0.9220 & 0.8569 \\
U-Net-Prep~\cite{Saifullah2024_icai3s} & 34,513,410 & 2004 & 0.8762 & 0.8891 \\
U-Net~\cite{Saifullah2023} & 34,513,410 & 1960 & 0.8759 & 0.8391 \\
\hline
\end{tabular}
\end{table}

Traditional U-Net-based methods, such as U-Net~\cite{Saifullah2023} and U-Net-Prep~\cite{Saifullah2024_icai3s}, are more computationally demanding, with parameter counts of 34,513,410 and execution times of 1960 and 2004 seconds, respectively. However, they achieve lower DSC (0.8759 and 0.8762) and IoU scores (0.8391 and 0.8891). In contrast, PSO-UNet’s optimized architecture achieves higher accuracy with fewer resources by dynamically optimizing hyperparameters such as the number of filters, kernel size, and learning rate. Its lightweight configuration (32 filters, kernel size 3, learning rate 0.0067) ensures high segmentation precision while minimizing computational overhead. 

This optimization allows PSO-UNet to deliver a solution that is both practical and efficient. Unlike resource-intensive traditional methods, PSO-UNet offers a lightweight solution without compromising accuracy. This makes it well-suited for clinical applications, as it achieves outstanding performance with reduced execution time and fewer parameters.

\section{Conclusion}
\label{conclusion}
This study introduces the PSO-UNet framework, which employs Particle Swarm Optimization (PSO) for dynamic hyperparameter optimization in U-Net-based medical image segmentation. The framework achieved state-of-the-art performance on the BraTS 2021 and Figshare datasets with Dice Similarity Coefficients (DSC) of 0.9578 and 0.9523, and Intersection over Union (IoU) scores of 0.9194 and 0.9097. With just 7.8 million parameters and an execution time of 906 seconds, PSO-UNet outperforms several benchmark methods in accuracy and computational efficiency, demonstrating its adaptability to diverse imaging modalities and tumor classes. Its lightweight design and dynamic optimization make it well-suited for clinical applications and resource-constrained environments, with the potential for further enhancement through attention mechanisms, hybrid optimization, and advanced preprocessing to handle complex datasets. The PSO-UNet sets a new approach for efficient and accurate medical image segmentation, offering a robust foundation for future advancements in medical imaging technologies.

\begin{acks}
Research funding was provided by AGH University of Krakow (Program ``Excellence initiative -- research university''), Polish Ministry of Science and Higher Education funds assigned to AGH University of Krakow, and ACK Cyfronet AGH (Grant no. PLG/2024/017503).
\end{acks}

\bibliographystyle{ACM-Reference-Format}
\bibliography{bibliography}










\end{document}